\documentstyle{mn}

\newif\ifAMStwofonts

\def \SAIT #1 #2 {{\em Mem.\ Soc.\ Astron.\ It.\/} {\bf #1}, #2}
\def \MESS #1 #2 {{\em The Messenger\/} {\bf #1}, #2}
\def \ASTRNACH #1 #2 {{\em Astron. Nach.\/} {\bf #1}, #2}
\def \AAP #1 #2 {{\em Astron. Astrophys.\/} {\bf #1}, #2}
\def \AAL #1 #2 {{\em Astron. Astrophys. Lett.\/} {\bf #1}, L#2}
\def \AAR #1 #2 {{\em Astron. Astrophys. Rev.\/} {\bf #1}, #2}
\def \AAS #1 #2 {{\em Astron. Astrophys. Suppl. Ser.\/} {\bf #1}, #2}
\def \AJ #1 #2 {{\em Astron. J.\/} {\bf #1}, #2}
\def \ANNREV #1 #2 {{\em Ann. Rev. Astron. Astrophys.\/} {\bf #1}, #2}
\def \APJ #1 #2 {{\em Astrophys. J.\/} {\bf #1}, #2}
\def \APJL #1 #2 {{\em Astrophys. J. Lett.\/} {\bf #1}, L#2}
\def \APJS #1 #2 {{\em Astrophys. J. Suppl.\/} {\bf #1}, #2}
\def \APSS #1 #2 {{\em Astrophys. Space Sci.\/} {\bf #1}, #2}
\def \ASR #1 #2 {{\em Adv. Space Res.\/} {\bf #1}, #2}
\def \BAIC #1 #2 {{\em Bull. Astron. Inst. Czechosl.\/} {\bf #1}, #2}
\def \JSQRT #1 #2 {{\em J. Quant. Spectrosc. Radiat. Transfer\/} {\bf #1}, #2}
\def \MN #1 #2 {{\em Mon. Not. R. Astr. Soc.\/} {\bf #1}, #2}
\def \MEM #1 #2 {{\em Mem. R. Astr. Soc.\/} {\bf #1}, #2}
\def \PLR #1 #2 {{\em Phys. Lett. Rev.\/} {\bf #1}, #2}
\def \PASJ #1 #2 {{\em Publ. Astron. Soc. Japan\/} {\bf #1}, #2}
\def \PASP #1 #2 {{\em Publ. Astr. Soc. Pacific\/} {\bf #1}, #2}
\def \NAT #1 #2 {{\em Nature\/} {\bf #1}, #2}
\def\strutdepth{\dp\strutbox}
\def\marginalstar{\strut\vadjust{\kern-\strutdepth\specialstar}}
\def\specialstar{\vtop to \strutdepth{\baselineskip\strutdepth\vss\llap{$\bigtriangleup\mskip-12mu\bigtriangledown$ }\null}}
\def \nm {N_{\mathrm{m}}}
\def \nm0 {N_{\mathrm{m,0}}}
\def \na {N_{\mathrm{a}}}
\def \na0 {N_{\mathrm{a,0}}}
\def \sm {\sigma_{\mathrm{m}}}

\def \BGE {\begin{equation}}
\def \EDE {\end{equation}}

\ifoldfss
  \ifCUPmtlplainloaded \else
    \NewTextAlphabet{textbfit} {cmbxti10} {}
    \NewTextAlphabet{textbfss} {cmssbx10} {}
    \NewMathAlphabet{mathbfit} {cmbxti10} {} 
    \NewMathAlphabet{mathbfss} {cmssbx10} {} 
  \fi
  \ifAMStwofonts
    \ifCUPmtlplainloaded \else
      \NewSymbolFont{upmath} {eurm10}
      \NewSymbolFont{AMSa} {msam10}
      \NewMathSymbol{\upi}     {0}{upmath}{19}
      \NewMathSymbol{\umu}     {0}{upmath}{16}
      \NewMathSymbol{\upartial}{0}{upmath}{40}
      \NewMathSymbol{\leqslant}{3}{AMSa}{36}
      \NewMathSymbol{\geqslant}{3}{AMSa}{3E}

    \fi
  \fi
\fi 

\ifnfssone
  \newmathalphabet{\mathit}
  \addtoversion{normal}{\mathit}{cmr}{m}{it}
  \addtoversion{bold}{\mathit}{cmr}{bx}{it}
  \newmathalphabet{\mathbfit} 
  \addtoversion{normal}{\mathbfit}{cmr}{bx}{it}
  \addtoversion{bold}{\mathbfit}{cmr}{bx}{it}
  \newmathalphabet{\mathbfss} 
  \addtoversion{normal}{\mathbfss}{cmss}{bx}{n}
  \addtoversion{bold}{\mathbfss}{cmss}{bx}{n}
  \ifAMStwofonts
    \ifCUPmtlplainloaded \else
      \UseAMStwoboldmath
      \makeatletter
      \new@mathgroup\upmath@group
      \define@mathgroup\mv@normal\upmath@group{eur}{m}{n}
      \define@mathgroup\mv@bold\upmath@group{eur}{b}{n}
      \edef\UPM{\hexnumber\upmath@group}
      \new@mathgroup\amsa@group
      \define@mathgroup\mv@normal\amsa@group{msa}{m}{n}
      \define@mathgroup\mv@bold\amsa@group{msa}{m}{n}
      \edef\AMSa{\hexnumber\amsa@group}
      \makeatother
      \mathchardef\upi="0\UPM19
      \mathchardef\umu="0\UPM16
      \mathchardef\upartial="0\UPM40
      \mathchardef\leqslant="3\AMSa36
      \mathchardef\geqslant="3\AMSa3E
    \fi
  \fi
\fi 

\ifnfsstwo
  \DeclareMathAlphabet{\mathbfit}{OT1}{cmr}{bx}{it}
  \SetMathAlphabet\mathbfit{bold}{OT1}{cmr}{bx}{it}
  \DeclareMathAlphabet{\mathbfss}{OT1}{cmss}{bx}{n}
  \SetMathAlphabet\mathbfss{bold}{OT1}{cmss}{bx}{n}
  \ifAMStwofonts
    \ifCUPmtlplainloaded \else
      \DeclareSymbolFont{UPM}{U}{eur}{m}{n}
      \SetSymbolFont{UPM}{bold}{U}{eur}{b}{n}
      \DeclareSymbolFont{AMSa}{U}{msa}{m}{n}
      \DeclareMathSymbol{\upi}{0}{UPM}{"19}
      \DeclareMathSymbol{\umu}{0}{UPM}{"16}
      \DeclareMathSymbol{\upartial}{0}{UPM}{"40}
      \DeclareMathSymbol{\leqslant}{3}{AMSa}{"36}
      \DeclareMathSymbol{\geqslant}{3}{AMSa}{"3E}
    \fi
  \fi
\fi 

\ifCUPmtlplainloaded \else
  \ifAMStwofonts \else 
    \def\upi{\pi}
    \def\umu{\mu}
    \def\upartial{\partial}
  \fi
\fi
   \title[Naked eye star visibility and limiting magnitude mapped from DMSP-OLS satellite data]{Naked eye star visibility and limiting magnitude mapped from DMSP-OLS satellite data}

   \author[P. Cinzano,        
          F. Falchi and 
          C. D. Elvidge 
           ]{P. Cinzano$^{1}$\thanks{E-mail: cinzano@pd.astro.it},        
          F. Falchi$^{1}$ and
          C. D. Elvidge$^2$ \\
$^{1}$ Dipartimento di Astronomia, Universit\`a di Padova,
vicolo dell'Osservatorio 5,  I-35122 Padova, Italy\\
$^2$ Office of the Director, NOAA National Geophysical Data Center, 325 Broadway, Boulder CO 80303}

\date{Accepted 14 November 2000.
      Received 7 November 2000;
      in original form 7 August 2000}

\pagerange{\pageref{firstpage}--\pageref{lastpage}}
\pubyear{2000}

\input epsf.sty
\begin{document}

\maketitle

\label{firstpage}

   \begin{abstract}

We extend the method introduced by Cinzano et al. (2000a) to map the artificial sky brightness in large territories from DMSP satellite data, in order to map the naked eye star visibility and telescopic limiting magnitudes. For these purposes we take into account the altitude of each land area from GTOPO30 world elevation data, the natural sky brightness in the chosen sky direction, based on Garstang modelling,  the eye capability with naked eye or a telescope, based on the Schaefer (1990) and Garstang (2000b) approach, and the stellar extinction in the visual photometric band. For near zenith sky directions we also take into account screening by terrain elevation.
Maps of naked eye star visibility and telescopic limiting magnitudes are useful to quantify the capability of the population to perceive our Universe, to evaluate the future evolution, to make cross correlations with statistical parameters and to recognize  areas where astronomical observations or popularisation can still acceptably be made.
We present, as an application,  maps of naked eye star visibility and total sky brightness in V band in Europe at the zenith with a resolution of approximately 1 km.

\end{abstract}


%

\section{Introduction}

The recent availability of high spatial resolution radiance calibrated night-time satellite images of the Earth (Elvidge et al.\ 1999) allows one to obtain quantitative information on the upward light flux emitted from almost all countries around the World (e.g. Isobe \& Hamamura 1998) bypassing problems arising when using population data to estimate light pollution:  (i) census data are not available everywhere, (ii) they are not updated frequently, (iii) they are based on city lists and do not provide spatially explicit detail of the population geographical distribution, (iv) they do not well represent some polluting sources, like e.g. industrial areas, harbours and airports, (v)  the upward emission per capita of a given city can deviate from the average and geographic gradients could exist.

In the last 16 years Roy Garstang has been carrying on a strong modelling effort (Garstang 1984, 1986, 1987, 1988, 1989a, 1989b, 1989c, 1991a, 1991b, 1991c, 1992, 1993, 2000a)  to develop an accurate technique to evaluate the night sky brightness produced by the upward light flux. It avoids resorting to empirical or semi-empirical formulae  which do not allow detailed relations with the atmospheric conditions, choice of the direction of observation and accounting for Earth curvature,  even if they are of invaluable utility for simple estimates (e.g. Walker 1973; Treanor 1973; Berry 1976; Garstang 1991b).

Cinzano et al. (2000a, hereafter Paper I) applied Garstang models to DMSP satellite data to produce detailed maps of the artificial night sky brightness across large territories opening the way to a quantitative analysis at global scale of the entire Earth (Cinzano et al. in prep.) and, joined to an even more continuous observation of the Earth made by DMSP satellites, to the prediction of the future evolution (Cinzano et al. 2000b; Cinzano et al. in prep.).

Both a comprehensive study of the effects of the increase of light levels in the night environment over their natural condition produced by wasted light and the evaluation of the effectiveness of  laws, standard rules and ordinances  to protect the environment and the capability of mankind to perceive the universe, require more than maps of artificial sky brightness at sea level.
Such maps, being free from elevation's effects, are useful for a detailed knowledge and comparison of the pollution levels across large territories and the recognition of most polluted areas or more polluting cities. They are also useful for the identification of dark areas and potential observatory sites. However they allow only  in an approximate way a quantitative evaluation of the capability of the population to see the heavens by naked eye or by a telescope, the determination of the falling trend of the limiting magnitude, their cross correlation with statistical parameters, the determination of the visibility of astronomical phenomena, the recognition of the areas of a territory where the perception of the Universe is more endangered or where astronomical observations or popularisation can be still acceptably made.
In fact (i) the altitude of a site not only acts on the levels of sky glow but also has non-negligible effects on stellar extinction, (ii) the natural sky brightness needs to be accounted for when computing the total sky brightness in low polluted sites, (iii) the relation between total sky brightness and visual limiting magnitudes is not linear, being related to eye capability to see a point source towards a bright background.

Here we extend the method of Paper I in order to be able to map naked eye star visibility and telescopic limiting magnitudes across large territories. As in Paper I, we evaluate the upward light flux based on DMSP satellite data and compute the maps modelling the  light pollution propagation in the atmosphere with Garstang models. 
They assume Rayleigh scattering by molecules and Mie scattering by aerosols and take into account extinction along light paths and Earth curvature. 
In this paper we take into account the altitude of each land area from GTOPO30 elevation data, the natural sky brightness in the chosen sky direction, based on the Garstang (1989) models, the eye capability or telescopic limiting magnitudes based on Garstang (2000b) and Schaefer (1990) approach, the stellar extinction in the visual photometric band based on Snell \& Heiser (1968) and Garstang (1989) formulae.  For near zenith sky directions we also take into account mountain screening.

In section \ref{s4} we describe our improvements to the mapping technique. In section \ref{elev} we deal with input data, describing  GTOPO30 elevation data, updating the reduction of satellite radiance data, summarizing the atmospheric  model. In section \ref{comp} we present the maps of naked eye star visibility and total sky brightness in V band in Europe at  zenith with a resolution of approximately 1 km, we compare map predictions with measurements of sky brightness and limiting magnitude and we discuss the screening effects. Section \ref{s7} contains our conclusions.

\section{Mapping technique}
\label{s4}
\subsection{Artificial sky brightness}

The total artificial sky brightness in a given direction of the sky in  $(x',\,y')$ is:
\begin{equation}
\label{int1}
b(x',\,y')=\int\int e(x,y) f((x,y),(x',\,y'))~dx ~dy\, ,
\end{equation}
where $e(x,\,y)$ is the upward emission per unit area in $(x,\,y)$, $f((x,\,y),(x',\,y'))$ is the light pollution propagation function, i.e. the artificial sky brightness per unit of upward light emission produced by unit area in $(x,\,y)$ in the site in $(x',\,y')$. 
When upward light flux is obtained from satellite measurements, the territory is divided into land areas with the same positions and dimensions as projections on the Earth of the pixels of the satellite image and each land area is assumed to be a source of light pollution with an upward emission $e_{x,y}$ proportional to the  radiance measured in the corresponding pixel multiplied by the surface area (see eq. 35). The total artificial sky brightness at the centre of each area, given by the expression (\ref{int1}), becomes:
\begin{equation}
\label{sum1}
b_{i,j}=\sum_{h}\sum_{k}  e_{h,k} f((x_{i},y_{j}),(x_{h},y_{k}))\, ,
\end{equation} 
 for each pair $(i,j)$ and $(h,k)$, which are the positions of the observing site and the polluting area on the array.

In paper I the method of mapping artificial sky brightness has been applied  (i) computing brightness at sea level, (ii) assuming sources at sea level and  (iii) assuming that the upward emission function has the same shape everywhere. In this case the light pollution propagation function $f$ depends only on the distance between the site and the source, and on some details which are assumed to be the same everywhere, such as the shape of the emission function, the atmospheric distribution of molecules and aerosols, their optical characteristics in the chosen photometric band and the direction of the sky observed. Eq. (\ref{sum1}), however, it is not a convolution because the distance between pairs of points depends on the latitude in the used latitude/longitude projection. If assumption (i) is relaxed, but assumptions (ii) and (iii)  are retained, a reasonable computational speed can be still obtained evaluating once for each latitude the array $f(d_{i-h, j-k}, h_{m} )$, where $d_{i-h, j-k}$ is the distance between the site and the polluting areas and $m$ is an index which discretizes the altitude $h$ of the site. Both $d_{i-h, j-k}$ and $h_{m}$ are computed inside a reasonable range (e.g. a circle with 200km of radius and the altitude of the highest mountain).
Then, all $b_{i,j}$ at nearly the same latitude can be rapidly obtained from eq. \ref{sum1} interpolating the array $f(d_{i-h, j-k}, h_{m} )$ at the elevation $h(i, j)$ of each site.
If assumptions (ii) and (iii)  are relaxed it becomes necessary to evaluate $f((x_{i},y_{j}),(x_{h},y_{k}))$ for each pair of points so that the computations become slower and at the moment  can be applied only to small territories. For this reason, we maintained assumptions (ii) and (iii) when computing the maps of Europe in sec. \ref{results}. 

We obtained the propagation functions $f(d_{i-h, j-k}, h_{m} )$ or $f((x_{i},y_{j}),(x_{h},y_{k}))$ with models for the light propagation in the atmosphere based on the modelling technique introduced and developed by Garstang  (1986, 1987, 1988, 1989a, 1989b, 1989c, 1991a, 1991b, 1991c, 1992, 1993, 2000a)  and also applied by Cinzano (2000a, 2000b). 
The propagation function $f$, expressed as total flux per unit area of the telescope
per unit solid angle per unit total upward light emission, is obtained for each set of indexes integrating along the line of sight:
\begin{equation}
\label{gar2}
f =\! \int^{\infty}_{u_{0}} \! \! \! \! \!\! \! \left( \beta_{\mathrm{m}}(h)f_{\mathrm{m}}(\omega)\!+\!\beta_{\mathrm{a}}(h)f_{\mathrm{a}}(\omega) \right) (1\!+\!D_{\mathrm{S}} )  i(\psi,s)  \xi_{1}(u) du ,     
\end{equation} 
where $\beta_{\mathrm{m}}(h)$, $\beta_{\mathrm{a}}(h)$ are respectively the scattering cross
sections  of molecules and aerosols per unit volume at the  elevation $h(u)$ along the line of sight,  $f_{\mathrm{m}}(\omega)$, $f_{\mathrm{a}}(\omega)$ are their normalized angular scattering functions, $\xi_{1}(u)$ is the extinction  of the light along its path to the telescope and $i(\psi,s)$  is 
the direct illuminance per unit flux produced by each source on each infinitesimal volume of atmosphere along the line-of-sight of the observer.
The  scattering angle $\omega$, the emission angle $\psi$, the distance $s$ of the volume from the
source and the elevation $h$ of it, depend on the altitudes of the site and the source, their distance, the  zenith distance and the azimuth  of the line-of-sight,  and  the distance $u$ along the line of sight, through some geometry described for curved Earth and nonzero altitude in Garstang (1989, eqs. 4-11).
The $(1+D_{\mathrm{S}} )$  in eq. (\ref{gar2}) is a correction factor which take into account the illuminance due to light already scattered once from molecules and aerosols which can be evaluated with the approach of Treanor \cite{tre73} as extended by Garstang (1984, 1986, 1989). Details for curved Earth can be found in the last paper (eq. 23) as well a discussion about the error in neglecting third and higher order scattering which can be significant for optical thickness higher than about 0.5.
We refer the reader to Paper I and the cited papers for details.

We can account for screening effects setting the illuminance per unit flux in eq. (\ref{gar2}) to:
\begin{equation}\label{gar4}
i(\psi,s)=I(\psi) \xi_{2}/ s^{2}
\end{equation}
where there is no screening by Earth curvature or by terrain elevation and zero elsewhere. 
Here $I(\psi)$ is the normalized emission function giving the relative light flux per unit solid angle emitted by each land area at the zenith distance $\psi$, $s$ is the distance between the source and the considered infinitesimal volume of atmosphere and $\xi_{2}$ is the extinction along this path. 
We considered every land area as a point source located in its centre except when $i=h$, $j=k$ in which case we used a four points approximation (Abramowitz \& Stegun 1964).
Taking into account screening effects requires us to check for each point along the line of sight whether the source area is screened by terrain elevation or not, taking into account Earth curvature. This can be done by determining the position of the foot of the vertical of the considered point and then computing for every land area crossed by a line connecting this foot and the source area, the quantity $\cot \psi$, where $\psi$ is defined as in figure 
\begin{figure}
\epsfysize7cm 
\hspace{0.4cm}\epsfbox{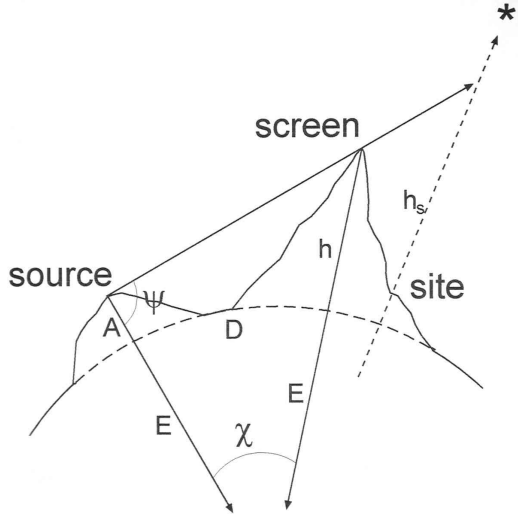} 
\caption[h]{Screening by terrain elevation.}
\label{figpsi}
\end{figure}
\ref{figpsi} and depends on the elevation $A$ of the land area, the distance $D$ of its center from the center of the source area and the Earth radius $E$:
\begin{equation}\label{psi1}
\cot  \psi = \frac{(A+E)-(h+E) \cos (D/E)}{(h+E) \sin (D/E)}
\end{equation}
Then we can determine  $X=\max (-\cot \psi)$  and from it the screening elevation $h_{\mathrm{s}}$:
\begin{equation}\label{psi2}
h_{\mathrm{s}} = \frac{A+E}{\cos (D^{\star}/E) - X \sin (D^{\star }/E)}-E
\end{equation}
where $D^{\star}$ is the distance between the source area and the foot of the vertical, and $h_{\mathrm{s}}$ is computed over the sea level.
The illuminance $i$ in eq. (\ref{gar2}) is set to zero when the elevation of the considered point is lower than the screening elevation.
For lines of sight pointing toward the zenith the evaluation of the screening elevation can be done once for each pair site-source. However, even in this faster case
the computational time required by the screening evaluation for each source area around each site is huge, so we accounted for screening effects due to terrain elevation only in the small maps of section \ref{screening} accounting in the other maps only for screening by Earth curvature as described by Garstang (1989, eq. 12-13).

\subsection{Natural sky brightness}
\label{natural}

The mapping of the total sky brightness requires the evaluation of the natural sky brightness under the atmosphere, i.e. as actually observed from the ground, in the direction of the line-of-sight.

We assumed as Roach \& Meinel (1955) and Garstang (1989) that natural sky brightness is produced by (i) light from a layer at infinity due mainly to integrated star light, diffused galactic light and zodiacal light, and (ii) light due to airglow emission from a van Rhijn layer at height of 130 km above the ground. The  first, $b_{\mathrm{s}}$, depends on the equatorial coordinates of the observation point and, for the zodiacal light, on the time $t$.  The second depends on the angle at which the layer is observed and on the layer brightness, $b_{\mathrm{vr}}$, which in turn depends on some factors like the geographical latitude $L$, the solar activity $S$ in the previous day, and the time $T$ after twilight.
Extending the Garstang (1989) approach, we  assumed that the natural sky brightness $b_{\mathrm{out}}$ outside the scattering and absorbing layers of the atmosphere, at zenith distance $z$ and azimuth $\omega$, is given by:
\begin{equation}\label{nat1}
b_{\mathrm{out}}(z, \omega, L, T, S)=  b_{\mathrm{s}}(\alpha, \delta, t) + \frac{b_{\mathrm{vr}}(L, T, S)}{ (1-0.96 \sin^{2} z )^{1/2}} 
\end{equation}
where $\alpha$ is the right ascension and $\delta$ the declination of the observed area of the sky which depends on the zenith distance $z$ and azimuth $\omega$ through the observation time $t$.
From Walker (1988) we know that the sky brightness decays to a nearly constant level after some hours from the astronomical twilight due probably to the recombination of ions excited during the day by the solar radiation, so we will refer the night brightness always to some hours after the twilight in order that the dependence on $T$ disappear.
The dependence of $b_{\mathrm{vr}}$ on the solar activity can be expressed in a very rough approximation as Cinzano (2000c), based on the measurements of Walker (1988), Cannon 1987 in Krisciunas et al. (1987), Krisciunas (1990, 1997):
\begin{equation}\label{nat2}
b_{\mathrm{vr}}=b_{\mathrm{vr},0}(L) \cdot 10^{ C \cos  (2 \pi\frac{t-t_{0}}{P})}
\end{equation}
where $b_{\mathrm{vr},0}$ is the value at mean solar activity, $P$ is the average period of the solar cycle,  $t_{0}$ is the epoch of a maximum, $t$ is the time from the epoch, and $C$ is a constant.
A good correlation was found by Walker (1988, see also Krisciunas 1997, eq. 5) between the sky brightness and the observed 10.7 cm solar radio flux for the day preceding the observations.
The value of $b_{\mathrm{s}}$ in a given direction on the sky, depending on the observation time, can be significant when maps are made to quantify the visibility of  an astronomical phenomenon. When the purpose is to evaluate the generic capability to see the stars, we can assume $b_{\mathrm{s}}$ constant and given by its average value at the considered latitude. 
In this case $b_{\mathrm{out}}$ does not depend on the azimuth.

The natural sky brightness $b_{\mathrm{nat}}$ observed at altitude $A$ over sea level is given by the sum of (i) the light directly coming to the observer $b_{\mathrm{d}}$, (ii) the light scattered by molecules $b_{\mathrm{m}}$ and (iii) the light scattered by aerosols $b_{\mathrm{a}}$:
\begin{equation}\label{nat3}
b_{\mathrm{nat}}=b_{\mathrm{d}}+ b_{\mathrm{m}} +b_{\mathrm{a}} 
\end{equation}
We computed these components using the model presented by Garstang (1989).

The directly transmitted light arrives at the observer after extinction along the line of sight (Garstang 1989):
\begin{equation}\label{nat4}
b_{\mathrm{d}}=b_{\mathrm{out}} \xi_{3}
\end{equation}
where $\xi_{3}$ is the atmospheric extinction in the path from infinity to the observer.

The light scattered by molecules by Rayleigh scattering for the atmospheric model of sec. \ref{atmod} is (Garstang 1989):
\begin{eqnarray}
b_{\mathrm{r}}\; \; \;\; &=&\! \!\frac{3(1+G)b_{\mathrm{out}} \beta_{\mathrm{m}}}{16\exp(cH) }  \int_{0}^{\infty}  i_{\mathrm{s}}(u,z) \exp(-ch) \xi_{4}du\\
i_{\mathrm{s}}(u,z)\! \! \! \! \!&=&\! \! \!\int_{0}^{1} \! \! \! \!  \left(2+2\mu^{2} \cos^{2} z+(1-\mu^{2}) \sin^{2} z\right) f(\mu) d\mu\\
f(\mu) \; &=&\! \!(\frac{b_{\mathrm{s}}}{b_{\mathrm{out}}}+\frac{b_{\mathrm{vr}}}{b_{\mathrm{out}}}(0.04+0.96 \mu^{2})^{-1/2}) \xi_{3} D_{\mathrm{s2}}   \\
D_{\mathrm{s2}}\; \; &=&\! \!1+11.11 K \beta_{\mathrm{m}} \exp(-cH) a^{-1}\mu^{-1} \exp(-ah)
\end{eqnarray}
where $u$ is the integration variable along the line of sight, $\beta_{\mathrm{m}}$ is the scattering cross section of molecules per unit volume,
$G$ is the ground reflectivity (assumed 15\%),
$c$ is the molecular inverse scale height,
$a$ is the aerosol inverse scale height,
$H$ is the elevation of the land,
$h$ is the elevation of the infinitesimal volume at $u$,
$A$ is the elevation of the site,
$\mu$ is an integration variable,
$K$ is the Garstang atmospheric clarity coefficient, 
$\xi_{4}$ is the atmospheric extinction in the path from the scattering volume to the observer. 

The light scattered by aerosols for the atmospheric model of sec. \ref{atmod} is (Garstang 1989):
\begin{eqnarray}
 b_{\mathrm{a}}\; \; &=& b_{\mathrm{out}} d_{\mathrm{a}} A_{\mathrm{M}} \xi_{3} D_{\mathrm{s3}}\\
d_{\mathrm{a}}\; \; &=&11.11 K \beta_{\mathrm{m}} \exp(-cH) a^{-1} \exp(-aA)\\
D_{\mathrm{s3}}&=&1+0.5 d_{\mathrm{a}} A_{\mathrm{M}}
\end{eqnarray}
where $A_{\mathrm{M}}$ is the airmass given by eq. (\ref{snell}).

Total sky brightness is $b_{\mathrm{T}\,i,j}=b_{i,j}+b_{\mathrm{nat}}$. We expressed  it  in photon radiance in ph cm$^{-2}$ s$^{-1}$ sr$^{-1}$  or in magnitudes per arcsec$^{2}$ with the Garstang (1986, 1989) relation:
\begin{equation}\label{phot2mag}
V_{i,j}=41.438-2.5 \log b_{\mathrm{T}\,i,j}
\end{equation}
We evaluated $b_{\mathrm{s}}$ and $b_{\mathrm{vr}}$ by fitting the predictions for the natural sky brightness at zenith with specific observations made in unpolluted sites after some hours from sunset and reduced $b_{\mathrm{vr}}$ approximately with eq. (\ref{nat2})  to average solar activity. We assumed $b_{\mathrm{s}}$ and $b_{\mathrm{vr},0}$ constant everywhere in Europe.

\subsection{Naked eye star visibility}
\label{eye}

We obtained the map of naked eye limiting magnitude as an array $m_{i,j}$, giving the limiting magnitude at each grid point,  from the array $V_{i,j}$, giving the total sky brightness, as described.

The illumination  $i$ in lux perceived by the eye from a source which is at the threshold 
of visibility to an observer when the brightness of observed background is $b_{\mathrm{obs}}$ in nanolambert and the stimulus size, i.e. the seeing disk diameter, is $\theta$ in arc minutes, was given by Garstang (2000b) based on measurements of Blackwell (1946) and Knoll, Tousey and Hulburt(1946):
\begin{eqnarray}\label{blackwell}
i_{1}& =& c_{1} (1 + k_{1} b^{1/2} )^{2} (1 + \alpha_{1} \theta^{2} + y_{1} b_{\mathrm{obs}}^{z_{1}} \theta^{2}) \\
i_{2} &=& c_{2} (1 + k_{2} b^{1/2} )^{2} (1 + \alpha_{2} \theta^{2} + y_{2} b_{\mathrm{obs}}^{z_{2}} \theta^{2}) \\
i \; \,  &=& i_{1} i_{2} /(i_{1} +i_{2} )     
\end{eqnarray}
with $c_{1} = 3.451 \times 10^{-9}$, $c_{2} = 4.276 \times 10^{-8}$, $k_{1} = 0.109$, $k_{2} = 1.51 \times 10^{-3}$, 
$y_{1} = 2.0 \times 10^{-5}$, y$_{2} = 1.29 \times  10^{-3}$, $z_{1} = 0.174$, $z_{2} = 0.0587$, $\alpha _{1} = 2.35 \times  10^{-4}$, $\alpha_{2} = 5.81 \times  10^{-3}$. 
The last equation is an artifact introduced by Garstang in order to  put together smoothly the two components $i_{1}$ and $i_{2}$, related respectively to the thresholds of  scotopic and photopic vision, obtaining the best fit with cited measurements. 

The observed background $b_{\mathrm{obs}}$ in eqs. (19-20) is related to the night sky background $b_{vis}$ in the visual band from (Garstang 2000b):
\begin{equation}
b_{\mathrm{obs}}=b_{\mathrm{vis}}/(F_{\mathrm{a}}F_{\mathrm{SC}}F_{\mathrm{cb}}) 
\end{equation}
where
$F_{\mathrm{a}}$ takes into account ratio between average pupil area of the Knoll, Tousey, Hulburt and Blackwell observers and the pupil area of the assumed observer,
$F_{\mathrm{SC}}$ takes into account the Stiles-Crawford effect,
$F_{\mathrm{cb}}$ allows for the difference in colour between the laboratory sources used in determining the
relationships between $i$ and $b$ and the night sky background. 
Sky brightness $b_{\mathrm{vis}}$ in visual band expressed in nanolambert can be obtained from sky brightness $V_{i,j}$ in V band, expressed in mag/arcsec$^{2}$,   inverting Garstang (1986, 1989) relation:
\begin{equation}\label{nano}
b_{\mathrm{vis}}=10^{-0.4 (V-26.346)}
\end{equation}

The illumination $i'$ produced over the atmosphere by a star at the threshold visibility are related to the threshold illuminations $i_{1}, i_{2}$ obtained from eqs. (19-20) for scotopic and photopic vision from (Garstang 2000b):
\begin{eqnarray}
i' \, &=& i'_{1} i'_{2} /(i'_{1} +i'_{2} ) \\
i'_{1}&=&F_{\mathrm{a},1}F_{\mathrm{SC},1}F_{\mathrm{cs},1}F_{\mathrm{e},1}F_{\mathrm{s},1}  i_{1}\\
i'_{2} &=&F_{\mathrm{a},2}F_{\mathrm{SC},2}F_{\mathrm{cs},2}F_{\mathrm{e},2}F_{\mathrm{s},2}  i_{2}
\end{eqnarray}
where $F_{\mathrm{a}}$ and $F_{\mathrm{SC}}$ are defined as before, $F_{\mathrm{cs}}$ allows for the difference in color between the laboratory sources and the observed star,
$F_{\mathrm{e}}$ allows for star light extinction in the terrestrial atmosphere, taking into account that star magnitudes are given {\it outside the atmosphere},
$F_{\mathrm{s}}$ allows for the acuity of any particular observer, defined so that $F_{\mathrm{s}}< 1$ leads to a lower threshold $i$ and therefore implies an eye
sensitivity higher than average due possibly to above average retinal sensitivity, observing experience or an above average eye pupil size.
The illumination $i'$ expressed in lux can be converted into magnitudes (Allen 1973 p. 197): 
\begin{equation}
m=-13.98-2.5\log i'  
\end{equation}

If $p_{0}$ is the pupil diameter used by the average of the Knoll, Tousey, Hulburt and Blackwell observers, who are assumed to have been age $A_{0}=23$, 
and $p$ is the pupil diameter of an observer aged $A$ when the sky background is $b_{\mathrm{obs}}$,  then $F_{a} = p_{0}^{2}/p^{2}$ from eq. (6) of Garstang (2000b) becomes:
\begin{eqnarray}
F_{\mathrm{a}}&=& \left( \frac {0.534\! -\! 0.00211 A_{0}\! -\! (0.236\! -\! 0.00127 A_{0})q}
{0.534\! -\! 0.00211 A\! -\! (0.236\! -\! 0.00127 A)q} \right)^{2}\\
q\; \; &=& \tanh (0.40 \log b_{\mathrm{obs}} - 2.20)
\end{eqnarray}
The Stiles-Crawford effect, due to the decreasing  efficiency in detecting photons with the distance from the center of the pupil, produces a non linear increase of sensibility when the eye pupil increases and  can be taken into account with equations from Schaefer (1990) modified as pointed out by Garstang (2000b) and Schaefer (1993). We neglected this effect which must be accounted for telescopic observations or whenever a larger precision is needed.

Differences in colour between the eye sensitivity curve and photometer sensitivity curve  used in determining  $i$ can be corrected with (Schaefer 1990):
\begin{eqnarray}
F_{\mathrm{cs},1}&=&10^{-0.4(1-(B-V)/2)}\\
F_{\mathrm{cs},2}&=&1
\end{eqnarray}
where we assumed as a typical star color index B$-$V = 0.7. Differences in colour between the laboratory background and the night sky background $F_{\mathrm{cb}}$ can be corrected with the same formula. The colour of night sky is difficult to evaluate when there is light pollution. Cinzano \& Stagni (2000) showed that the sky becomes redder far from sources. However  near predominant sources, like large cities, where the extinction is negligible and aerosol scattering is large, the colour index is related to the colour of the integrated lamp spectra.  We assumed here B$-$V = 0.7 on average but when emission spectra of each land area become available, it will be possible to obtain the colour index of the night sky point by point computing maps of total sky brightness separately for B and V bands like in Paper I.
Stellar extinction in the atmosphere $F_{\mathrm{e}}$ is computed from eq. (\ref{snell}) and from the V band vertical extinction (sec. \ref{atmod}) corrected approximately for the night vision as Schaefer (1990). 
In the computation of  $i'_{1}$ and $i'_{2}$ must be used respectively the correction factors for scotopic and photopic vision.
The reader is referred to Schaefer (1990) and Garstang (2000b) for an extensive discussion.

\subsection{Telescopic limiting magnitude}

We can also obtain maps of telescopic limiting magnitude for a given instrumental setup. This could be useful for amateurs observational campaigns. In this case we must replace the image size $\theta $ by $M \theta $ in eq. (\ref{blackwell}), where $M$ is the magnification of the
telescope. 

The observed background $b_{\mathrm{obs}}$ is related to the night sky background under the atmosphere  $b_{\mathrm{vis}}$ from (Garstang 2000b):
\begin{equation}
b_{\mathrm{obs}}=b_{\mathrm{vis}}/(F_{\mathrm{b}}F_{\mathrm{t}}F_{\mathrm{p}}F_{\mathrm{a}}F_{\mathrm{m}}F_{\mathrm{SC}}F_{\mathrm{c}})  
\end{equation}
where
$F_{\mathrm{a}}$ takes into account the ratio of the area of the telescope to that of the naked eye,
$F_{\mathrm{SC}}$ takes into account the Stiles-Crawford effect,
$F_{\mathrm{cb}}$ allows for the difference in colour between the laboratory sources used in determining the relationships between $i$ and $b$ and the night sky background, 
$F_{\mathrm{b}}$ takes into account that one eye is used in telescopic observations, while binocular vision was used in obtaining the relations between $i$ and $b$,
$F_{\mathrm{t}}$ allows for the loss of light in the telescope, $F_{t}$ being the reciprocal of the transmission $t$ through the telescope and eyepieces,
$F_{\mathrm{p}}$ allows for the loss of light if the telescope exit pupil is larger than the eye pupil,
$F_{\mathrm{m}}$ allows for the reduction of the sky brightness by the telescope magnification

The illuminations perceived from a star are related to the illuminations given by eqs. (19-20) for scotopic and photopic vision from (Garstang 2000b):
\begin{eqnarray}
i'_{1}&=&F_{\mathrm{a},1}F_{\mathrm{SC},1}F_{\mathrm{cs},1}F_{\mathrm{e},1}F_{\mathrm{s},1}F_{\mathrm{b},1}F_{\mathrm{t},1}F_{\mathrm{p},1}   i_{1}\\
i'_{2}&=&F_{\mathrm{a},2}F_{\mathrm{SC},2}F_{\mathrm{cs},2}F_{\mathrm{e},2}F_{\mathrm{s},2}F_{\mathrm{b},2}F_{\mathrm{t},2}F_{\mathrm{p},2}   i_{2}
\end{eqnarray}
where $F_{\mathrm{cs}}$, $F_{\mathrm{e}}$ and $F_{\mathrm{s}}$ has been already discussed in sec. \ref{eye}.

We refer the reader to Schaefer (1990) and Garstang (2000b) for the formulae and further discussions. Note that, as pointed out by the last author, Schaefer's $F_{\mathrm{r}}$ is not needed because  the image size was already included in eq. (\ref{blackwell}).

\section{Input data}
\label{elev}
\subsection{Altitude data}

As input elevation data we used GTOPO30, a global digital elevation model (DEM) by the U.S. Geological Survey's
EROS Data Center. Details  have been given by  Gesch et al. (1999).
This global data set covers the full extent of latitude and longitude  with an horizontal grid spacing of
30-arc seconds as does our composite satellite image.  From the global 16-bit DEM (21,600 rows by 43,200 columns),  provided as 16-bit signed integer data in a simple binary raster, we cut an array of 4800$\times$4800 pixels covering the same area as our satellite image.
The vertical units represent elevation in meters above mean sea level which
 ranges from -407 to 8,752 meters. We reassigned a value of zero to ocean
areas,  masked as "no data" with a value of -9999, and to altitudes under sea level.

GTOPO30 is based on data derived from 8 sources of elevation information,
including vector and raster data sets: (i)
Digital Terrain Elevation Data (DTED) a raster topographic data base with
a horizontal grid spacing of 3-arc seconds (approximately 90 meters) produced
by the National Imagery and Mapping Agency (NIMA); (ii)                              
Digital Chart of the World,  a vector cartographic data set based on
the 1:1,000,000-scale Operational Navigation Chart series products of NIMA; (iii)                                   
USGS 1-degree DEM's with an horizontal grid spacing of 3-arc seconds;    (iv)                                       
Army Map Service 1:1,000,000-scale paper maps  (AMS) digitized by Geographical Survey Institute (GSI) of Japan;   (v)                  
International Map of the World 1:1,000,000-scale  (IMW) digitized by GSI  for the Amazon basin;       (vi)
digitized Peru 1:1,000,000-scale map  to fill gaps in source data for South America;     (vii)                               
Manaaki Whenua Landcare Research  DEM with a 500-meter
horizontal grid spacing for New Zealand;  (viii)                                 
Antarctic Digital Database (ADD) under the auspices of the
Scientific Committee on Antarctic Research.

The absolute vertical accuracy  varies by location according to
the source data and
at the 90 percent confidence level is  
30 m for DTED,    
 160 m for  DCW,          
30  m   for USGS DEM,          
250   m for AMS maps,          
 50 m for IMW maps,   
500 m for Peru map,    
 15 m for N.Z. DEM and
 not available for ADD (Gesch et al. 1999).   
For many areas the relative accuracy is probably better than the estimated
absolute accuracy.

As discussed by Gesch et al. (1999), due to the nature of the raster
structure of the DEM, small islands in the ocean less than approximately
1 square kilometre are not represented. Nonetheless the error in assuming them at sea level is usually small because their limited size do not allow very high elevations.
Not all topographic features that one would expect to be resolved at 30-arc second grid 
spacing are represented but this grid spacing is appropriate for the areas derived
from higher resolution DEM's. 
Changes in detail of topographic information are evident at the
boundary between two sources, even if
the mosaicing techniques smooth the
transition areas.  
Artefacts due to the production method are plainly visible in some areas even if their magnitudes  in a local area are usually
well within the estimated accuracy for the source.
Some production artefacts are also present in the areas derived from the
vector sources.  Small artificial mounds and depressions may be present in
localized areas, particularly where steep topography is adjacent to
relatively level areas, and the data were sparse.  Additionally,
a "stair step" (or terracing) effect may be seen in profiles of some areas,
where the transition between contour line elevations does not slope
constantly across the area but instead is covered by a flat area with
sharper changes in slope at the locations of the contour lines.

\subsection{Upward flux data}
\label{ss}

Upward flux data have been obtained from the Operational Linescan System (OLS) carried by the U.S. Air Force Defense Meteorological Satellite Program (DMSP) satellites. 
This is an oscillating scan radiometer  with low-light visible and thermal infrared (TIR) imaging capabilities. 
At night the OLS, carrying a 20 cm reflector telescope, uses a Photo Multiplier Tube (PMT) to intensify the visible band signal which have a broad spectral response covering the range for primary emissions from the most widely used lamps for external lighting. 
Details are described in Paper I, Lieske \cite{lieske}, Elvidge et al. (1997a, 1997b, 1997c, 1999).

The collection of special DMSP data, used in Paper I and here to assemble a cloud-free composite image calibrated to top-of-atmosphere radiances, has been obtained  after a special requests to the Air Force made by the U.S. Department of Commerce, NOAA National Geophysical Data Center (NGDC), which serves as the archive for the DMSP and develops night time lights processing algorithms and products. 
The primary reduction steps include (see Paper I and Elvidge et al.\ 1999):
\\1) acquisition of special OLS-PMT data at a number of reduced gain settings (24dB, 40dB, 50dB) to avoid saturation on major urban centres and, in the same time,
overcome PMT dynamic range limitation (Our request was granted for the darkest nights of lunar cycles in March 1996 and January-February 1997.).  On board algorithms which adjust the visible band gain  were disabled. 
\\2) establishment of a reference grid with finer spatial resolution than the input imagery;
\\3) identification of the cloud free section of each orbit based on OLS-TIR data;
\\4) identification of lights, removal of noise and solar glare, cleaning of defective scan lines and cosmic rays;
\\5) projection of the lights from cloud-free areas from each orbit into the reference grid;
\\6) calibration to radiance units using prior to launch calibration of digital numbers for given input telescope illuminance and  gain settings in simulated space conditions;
\\7) tallying of the total number of light detections in each grid cell and calculation of the average radiance value;
\\8) filtering images based on frequency of detection to remove ephemeral events;
\\9) transformation  into latitude/longitude projection with 30''$\times$30'' pixel size; 
\\10) cutting of  the requested portion  of the final image (we used an image of  4800$\times$4800 pixel corresponding to 40\degr$\times$40\degr starting approximately at longitude 10\degr 30' west and latitude 72\degr north). 

We improved map predictions by applying to the composite satellite image a mild deconvolution with the Lucy-Richardson algorithm. In fact, (i)  the effective instantaneous field of view  is larger (2.2 km at nadir to 5.4 km at the scan edges) than pixel-to-pixel ground sample distance (GSD) maintained by the along-track OLS sinusoidal scan and the electronic sampling of the signal from the individual scan lines (0.56 km), (ii) most of the data received by NGDC has been "smoothed" by on-board averaging of 5 $\times$ 5 pixel blocks, yielding data with a GSD of 2.8 km, and (iii) data of more orbits have been tallied.   A mild deconvolution allows partial recovery of the smeared radiance and allows better predictions for sites near strong sources like cities where spreading in distribution of upward emission could have an effect on map results. The point spread function has been obtained searching for isolated nearly-point sources and the deconvolution has been applied to smaller subimages. We plan in future analysis to download from the satellite the original high-resolution data in order to properly deconvolve single orbit data before tallying.
We also plan in future reductions to correct data for atmospheric extinction before tallying and to use only data from areas not very far from nadir in order to avoid effects of the shape of the upward emission function when changing the observation angle. As showed in Paper I, however, this is only a second order effect due to opposite contributions of extinction and emission function shape.

Calibrated upward flux measurements can be obtained based on pre-fly irradiance calibration of OLS PMT as described in Paper I.
If $\overline{r}$ is the energetic radiance in $\left[\mathrm{10^{10} W cm^{-2} sr^{-1} }\right]$ measured by the OLS-PMT,  the upward light flux $e$ in $\left[\mathrm{V\; band \; photons \; s^{-1} }\right]$ is given by eq. (28), (29) and (30) of Paper I:
\begin{equation}
\label{prefcab}
e  =\overline{r}\frac{\int^{\infty}_{0} T_{\mathrm{V}} \,I_{\lambda}   \lambda  d\lambda}
{\int^{\infty}_{0} T_{\mathrm{P\! M\! T}}\, I_{\lambda} \lambda d\lambda}\,
\frac{<\lambda>}
{hc}\,  \frac{10^{0.4 \Delta m}\cos(l) }{I(\psi\! =\! 0)} \left( \frac{2 \pi \Delta x R_{\mathrm{T}}}{1.296\! \cdot\!  10^{6}}  \right)^{2}\! ,
\end{equation}
where $I(\psi)$ is the normalized upward emission function as defined in paper I, $l$ the latitude of the land area, $R_{\mathrm{T}}$  the average Earth radius in km and $\Delta x$ the pixel size in arcsec, 
$T_{\mathrm{V}}$ and $T_{\mathrm{P\! M\! T}}$ the sensitivity curves respectively of V band and PMT detector, $I_{\lambda}$ the energy spectrum of the emission from the chosen land area,  $h$ the Planck constant, $c$ the velocity of light and $<\lambda>$ the effective wavelength of the combination of the sensitivity curves of the PMT and the calibration source. 
We assumed  an average vertical extinction $\Delta m=0.33$ mag V for all measured land areas neglecting their elevation and solved the integrals in eq. (\ref{prefcab}) constructing, as in Paper I, an approximate synthetic spectrum for a typical night-time lighting roughly assuming that 50 per cent of the total emitted power be produced by High Pressure Sodium (HPS) lamps (SON standard) and 50 per cent by Hg vapour lamps (HQL).

We assumed that all land areas have on average the same normalized emission function, given by the parametric representation of Garstang \cite{g86}  in eq. (15) of  Paper I. This has been tested by Garstang and by Cinzano (2000a) with many comparisons between model predictions and measurements and in Paper I by studying in a single orbit satellite image the relation between the upward flux per unit solid angle per inhabitant of a large number of cities and their distance from the satellite nadir, which is related to the emission angle $\psi$. See Paper I for a detailed discussion.

\subsection{Atmospheric data and stellar extinction}
\label{atmod}

In order to compute extinctions along the light paths and scatterings, we need atmospheric data or an atmospheric model for the considered territory. In principle what we need is a set of functions
 giving, for each triplet of longitude, latitude and elevation $(x, y, h)$, the molecular and aerosol scattering coefficients per unit volume of atmosphere $\beta_{\mathrm{m}}(x,y,h)$ and $\beta_{\mathrm{a}}(x,y,h)$, and the aerosol angular scattering function $f_{\mathrm{a}}(\omega, x, y, h)$.
The molecular angular scattering function $f_{\mathrm{m}}(\omega)$ is known because it is Rayleigh scattering. If discretised in arrays, they should preferably have the same grid spacing of our upward flux and elevation data.
The atmospheric data or model would need to refer to conditions for {\it typical} clean nights at every point and contain any other information on denser aerosol layers, volcanic dust and the Ozone layer. At the moment this is not at our disposal. 
Moreover, applying typical condition in every land area we risk mixing effects due to light pollution with effects due to gradients of atmospheric conditions in typical nights. So we  applied the same standard atmospheric model everywhere, neglecting geographical gradients and local particularities as in Paper I.
Here we resume the simple atmospheric model.

1) We assumed the molecular atmosphere in hydrostatic equilibrium under the gravitational force as Garstang \cite{g86} with an inverse scale height $c=\mathrm{0.104\,    km^{-1}}$, a molecular density at sea level $N_{\mathrm{m,0}} =\mathrm{2.55\times 10^{19} cm^{-3}}$ and a constant integrated  Rayleigh scattering cross section  in V band $\sm=4.6 \times10^{-27}$ cm$^{2}$ molecule$^{-1}$. The scattering cross section per unit volume is $\beta_{\mathrm{m,0}}=N_{\mathrm{m,0}} \sigma_{\mathrm{m}}$. Note 
that in sec. 4.2 of Paper I  the letters "B" and "V" of the photometric bands of the Rayleigh cross sections have been exchanged and that the units are $\left[\mathrm{cm^{2}} \; \mathrm{molecule^{-1}} \right]$.

2) We assumed  the atmospheric haze aerosols number density decreasing exponentially with the altitude as Garstang \cite{g86} and Joseph et al. \cite{joseph} neglecting
the presence of sporadic denser aerosol layers, volcanic dust and the Ozone layer, studied by Garstang (1991a, 1991c). As Garstang \cite{g86}, the inverse scale height of aerosols was assumed to be  $a=0.657+0.059 K$. Aerosol content was given using the Garstang atmospheric clarity parameter $K$ which measures the relative importance of aerosol and molecules for scattering light in V band:
\begin{equation}\label{kappa}
K=\frac{\beta_{\mathrm{a,0}} }{\beta_{\mathrm{m,0}}  11.11 \mathrm{e}^{-cH}}\,  , 
\end{equation}
where $H$ is the altitude of the ground level above sea level.
The typical normalized angular scattering function for atmospheric haze aerosols was assumed to be given by the  function tabulated by Mc Clatchey et al. \cite{mccl} as interpolated by  Garstang \cite{g91a}. 

The stellar extinction at zenith in magnitudes for a site at altitude $A$   is given for this atmospheric model by Garstang (1986):
\begin{equation}\label{vertext}
\Delta m= 1.0857 \beta_{\mathrm{m,0}}  e^{-cH}\left(  \frac{e^{-cA}}{c}+ 
 \frac{11.778 Ke^{-aA}}{a} \right),
\end{equation}
The stellar extinction at zenith distance $z$ can be obtained from the 
 air mass $A_{\mathrm{M}}$ given by Snell \& Heiser (1968):
\begin{equation}
\label{snell}
A_{\mathrm{M}}=\sec z - g (\sec z - 1)^{2}
\end{equation}
with $g=0.010$ as chosen by Garstang (1986) to reproduce the table I, column 3 of Allen (1973) to better than 0.1 air masses at zenith distance 85\degr.
The atmospheric extinctions   $\xi_{1}$, $\xi_{2}$, $\xi_{3}$, $\xi_{4}$ of sec. \ref{s4} are given for this atmospheric model by Garstang (1989,  eqs. 18-22).
We can associate the atmospheric conditions with other observable quantities like
the horizontal visibility (Garstang 1989, eq. 38), the optical thickness $\tau$ (Garstang 1986, eq. 22) and the Linke turbidity factor for total solar radiation  (Garstang 1988).

\section{Results}
\label{comp}

\subsection{Maps of total sky brightness and star visibility}
\label{s6}
\label{results}
We present as an application the map of zenith night sky brightness and naked eye star visibility in Europe.

\begin{figure}
\epsfysize=19cm 
\hspace{0.4cm}\epsfbox{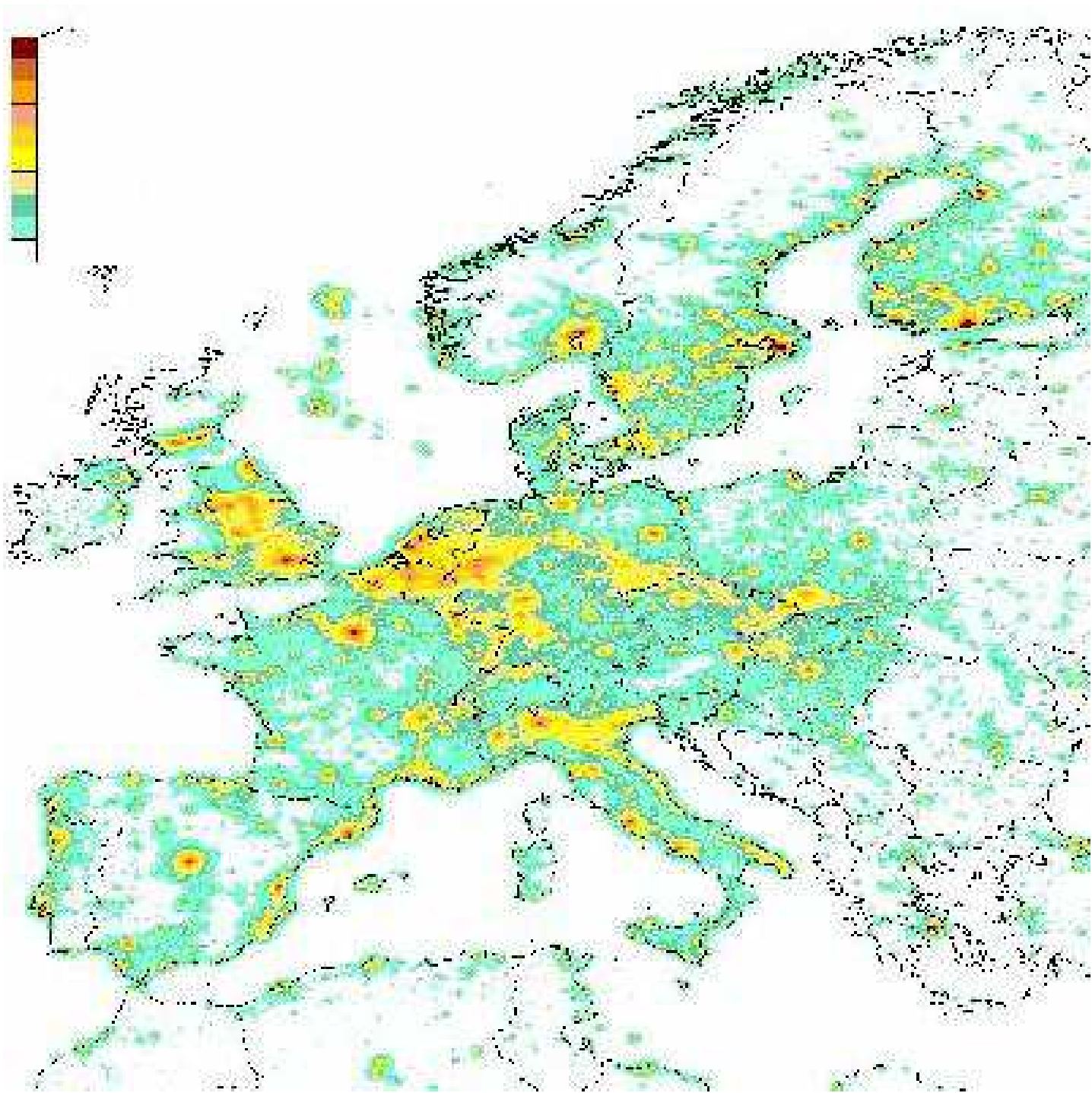} 
\caption[h]{Total night sky brightness  in Europe in V band for aerosol content parameter $K=1$.}
\label{res1}
\end{figure}

Fig. \ref{res1} shows the total sky brightness at the zenith in V photometric astronomical band (Johnson 1955).  
Colour levels from brown to white correspond to total sky brightness  of: $<$17.5, 17.5-18, 18-18.5, 18.5-19, 19-19.5, 19.5-20, 20-20.5, 20.5-21, 21-21.5,  $>$21.5 V mag/arcsec$^{2}$. The map was computed for clean atmosphere with aerosol clarity $K=1$, corresponding to a vertical extinction in V band of $\Delta m =0.33$ mag at sea level, $\Delta m =0.21$ mag at 1000m o.s.l., $\Delta m =0.15 $ mag at 2000m o.s.l.,  horizontal visibility at sea level $\Delta x=26$ km, optical depth $\tau=0.3$ so double scattering approximation is adequate. Each pixel is 30''$\times$30'' in size in longitude/latitude projection.  
Computation of the natural sky brightness is based on measurements at Isola del Giglio (Italy), a quite dark site according to the World Atlas of Artificial Sky Brightness (Cinzano et al. in prep.), giving  $V=21.74\pm0.06$ mag arcsec$^{-2}$ in V band in 1999 at 200m o.s.l.. At that time average solar activity was reported. The map was rescaled from 1996-1997 to 1998-1999 adding $\Delta m = 0.28$ mag/arcsec$^{2}$ to the total sky brightness obtained from OLS-PMT pre-fly radiance calibration. This correction 
was obtained fitting a straight line $y=\Delta m+x$ to the observed versus predicted data points of sec. \ref{calibtotbri}. The difference $\Delta m=0.28\pm0.10$ mag/arcsec$^{2}$ is possibly due to the growth of light pollution and agree within errorbars with Cinzano (2000c).

\newpage
\onecolumn
\begin{figure}
\epsfysize=19cm 
\hspace{0.4cm}\epsfbox{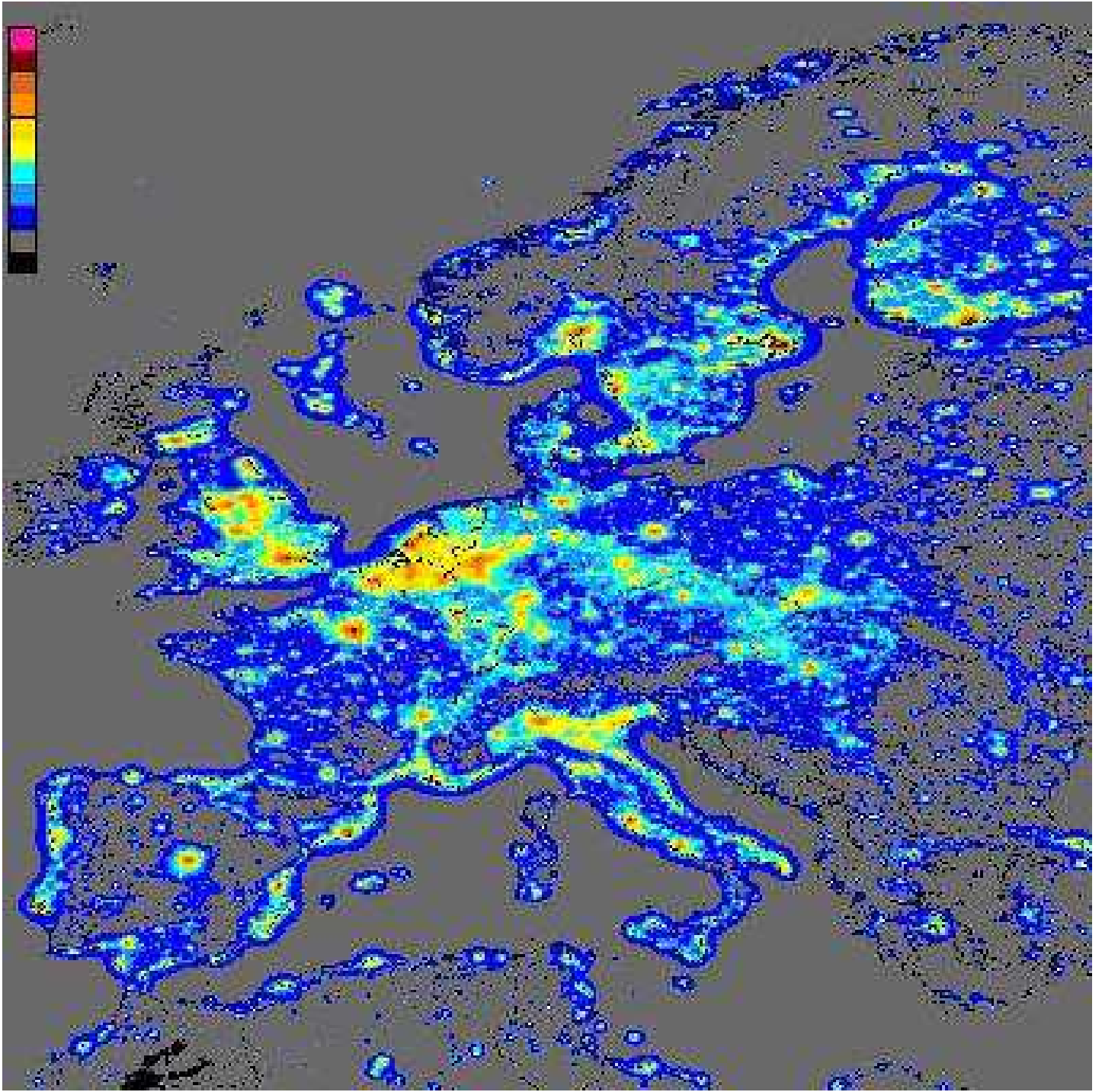} 
\caption[h]{Naked eye limiting magnitude  in Europe  for aerosol content parameter $K=1$.}
\label{res3}
\end{figure}
\twocolumn
\newpage

The limiting magnitude is a statistical concept (Blackwell 1946; Schaefer 1990).  A number of random factors affect eye measurements like the individual eye
sensitivity,
 the difference from the average eye pupil size, the capability to use averted vision, the experience which make an observer confident of a detection at a probability level different from the others (Schaefer 1990 reports a difference of 1 magnitude from 10\% to 90\% detection probability, corresponding
respectively to the fainter suspected star and the fainter surely visible star), the lenghts of time for which the field has been observed (Schaefer 1990 reports roughly half a magnitude from 6-seconds to 60-seconds observation). Fig. \ref{res3} shows the centre of the gaussian distribution of the naked eye limiting magnitude in Europe at the zenith obtained from the map of fig. \ref{res1}.  
Colour levels from pink to black correspond to: $<$3.75, 3.75-4.00, 4.00-4.25, 4.25-4.50, 4.50-4.75, 4.75-5.00, 5.00-5.25, 5.25-5.50, 5.50-5.75, 5.75-6.00, $>$6.00 V magnitudes. 
Limiting magnitudes are computed for observers of average experience and capability $F_{\mathrm{s}}=1$, aged 40 years, with the eyes adapted to the dark, observing with both eyes. Observer experience can be accounted with eq. (20) of Schaefer (1990).

Original maps are 4800$\times$4800 pixel images saved  in 16-bit standard {\sc fits} format  with {\sc fitsio} Fortran-77 routines developed by HEASARC at the NASA/GSFC. They have been analysed with {\sc ftools 4.2} analysis package by HEASARC and with {\sc Quantum Image 3.6} by Aragon Systems.   
Readers should consider the distinction between grid spacing and resolution.
The resolution of the maps, depending on results from an integration over a large zone, is greater than resolution of the original deconvolved images and is generally of the order of the distance between two pixel centres (30''$\times$30'', i.e. less than 1 km). Country boundaries are approximate.

\subsection{ Comparison with total sky brightness measurements} 
\label{calibtotbri}

\begin{figure}
\epsfysize=8.5cm 
\hspace{-0.5cm}\epsfbox{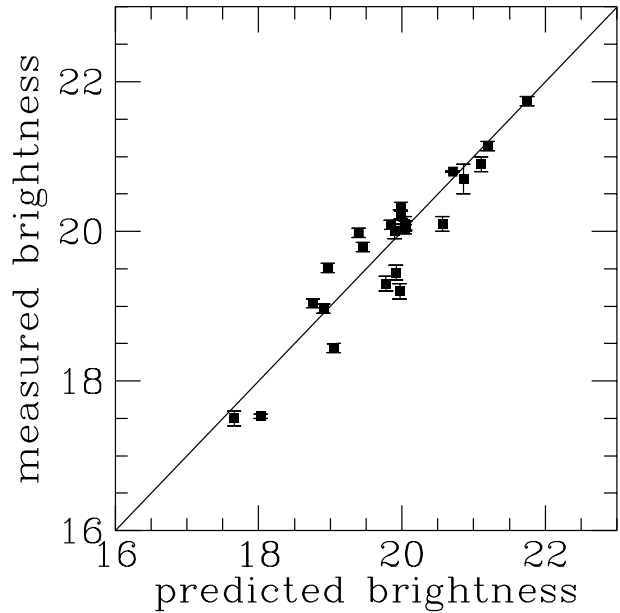} 
\caption[h]{Measurements of total sky brightness versus map predictions in V band. The straight line is the linear regression.}
\label{calv}
\end{figure}

Fig. \ref{calv} shows a comparison between predictions of total night sky brightness and measurements in Europe in V band in 1998 and in 1999. 
A detailed comparison  requires measurements taken  (i) at a large number of  sites, (ii) in nights with the same vertical extinction and horizontal visibility assumed in the map  computation, (iii) in many similar nights in order to smooth atmospheric fluctuations by averaging, (iv)  under  the atmosphere, i.e. as actually observed from the ground without any extinction correction applied, (v) in the same period in which the satellite image was taken in order to minimize uncertainties given by the fast growth rate of artificial sky brightness, (vi) in the same photometric band for which maps are computed, (vii) with accurate geographical positions.  
As in Paper I, due to the scarcity of measurements of sky brightness associated to measurements of extinction, or to any other index of the atmospheric aerosol content, we used all available measurements taken in clean or photometric nights even if extinction was not available or not exactly the required one (Catanzaro \& Catalano 2000; Cinzano 2000a; Della Prugna 1999; Favero et al.\ 2000; Piersimoni, Di Paolantonio \& Brocato 2000; Poretti \& Scardia 2000; Zitelli 2000, Falchi 1998). Differently from  Paper I we did not need to subtract an assumed natural sky brightness from measurements to obtain artificial sky brightness. Errorbars indicate measurement errors which are much smaller than the uncertainties produced by fluctuations in atmospheric conditions which are unknown. Shifts in measurements obtained with different instrumental setups could also arise, as pointed out in Paper I.  
The best fit of a straight line  $y=a+bx$ to the data points (solid line in fig. \ref{calv}), assuming unknown uncertainties,  gives $b=0.99\pm0.08$ and $a=-0.21\pm1.58$ mag/arcsec$^{2}$.
The sigma derived from the chi-square assuming that our predictions fit well, $\sigma =  0.35$ mag/arcsec$^{2}$, give an estimate of the uncertainty of our predictions at a site. When a large number of measurements of sky brightness together with their stellar extinction will be available, a more precise evaluation of the uncertainty become possible. A worldwide CCD measurement campaign of both sky brightness and stellar extinction has been organized by the International Dark-Sky Association (Cinzano \& Falchi 2000). 

\subsection{Comparison with naked eye limiting magnitude measurements} 
\label{caliblimit}

We compared our predictions of naked eye limiting magnitude with observations taken in Europe in the period 1996-1999. 
Data have been obtained from preliminary results of the measurement campaigns set up by a number of organizations: (i) Dataset A: Operation 'Atlas 1996', Comit\'e National Pour la Protection du Ciel Nocturne, France (Corp 1998); (ii) Dataset B: 'Gli studenti fanno vedere le stelle', Ministero della Pubblica Istruzione - Unione Astrofili Italiani - Legambiente, Italy (Corbo 2000a,b);  (iii) Dataset C: Astronomy On-line 'Light Pollution Project', European Southern Observatory (Haenel 1999); (iv) Dataset D: 'CCD Amateurs Measurements of Night Sky Brightness', International Dark-Sky Association (IDA) - Italian Section (Falchi priv. comm.).  
\begin{figure}
\epsfysize=14cm 
\hspace{-0.85cm}\epsfbox{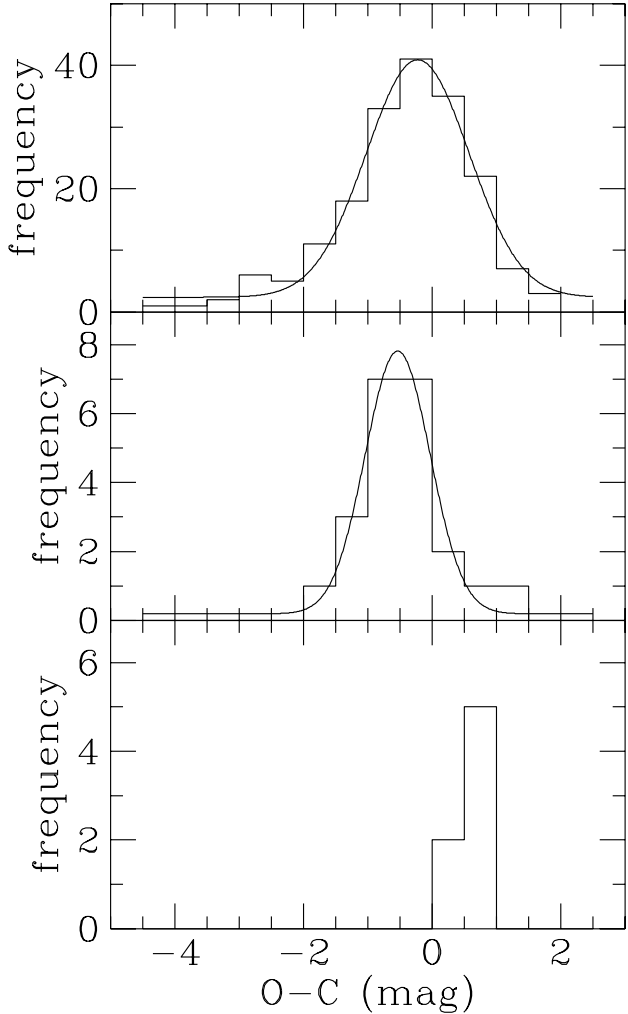} 
\caption[h]{Distribution of the O-C residuals for naked eye limiting magnitudes in dataset A (upper panel), datasets B and C (middle panel), dataset D (lower panel). Best fitting gaussians are also shown in the two upper panels.}
\label{calb1}
\end{figure}
\begin{figure}
\epsfysize=8.5cm 
\hspace{-0.5cm}\epsfbox{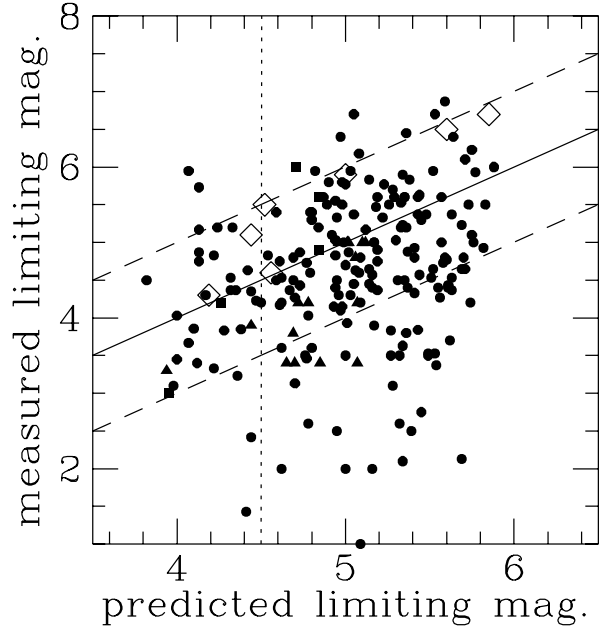} 
\caption[h]{Measurements of naked eye limiting magnitude versus map predictions for dataset A (dots), datasets B and C (triangles) and dataset D (open diamonds). The dotted line separates approximately the areas with prominent photopic vision (left) and scotopic vision (right). Solid line shows O-C=0 and the dashed lines show deviations of $\pm1$ mag.}
\label{calb2}
\end{figure}
A detailed comparison between map predictions of naked eye limiting magnitudes and visual estimates requires observations made (i) at a large number of  sites,  (ii) by a large number of observers in each site in order to have a statistical treatment of eye capabilities, (iii) in nights with the same vertical extinction and horizontal visibility assumed in the map  computation, (iv) in many similar nights in order to smooth atmospheric fluctuations by averaging,  (v) in the same period in which the satellite image was taken, (vii) with accurate geographical positions (better than 15'').   
The number of measurements  from each site at our disposal was too little to allow a statistical analysis site by site. Moreover, excluding the IDA data, the measurements have been taken without contemporary photometrical measurements of extinction so that the effects of random atmospheric content add further uncertainty. Uncertainties in geographical position could also exist. 
Systematic errors could also arise if the majority of observers detected the faint suspected star and not the fainter surely seen. We  excluded measurements for which the observer reported the presence of the moon, unclean sky, large light installations in the nearby or for which we were unable to determinate the geographical position. 

Fig.  \ref{calb1} shows the distribution of the observed minus calculated limiting magnitudes. The upper 
panel shows dataset A (180 sites) made for 50\% by active amateurs, for 25\% by very experienced amateurs and for 25\%  by beginners. A gaussian give a very good fit with $R^{2}=0.986$, a shift of the centre of only $\Delta x =-0.22\pm0.03$ mag and a dispersion $\sigma=0.79\pm0.04$ mag. Schaefer (1990) compared 314 visual observations with his model of limiting magnitude obtaining a nearly-gaussian distribution of errors with an HWHM of 0.75 mag and a shift of the centre of -0.24 mag. The middle panel show  measurements coming from datasets B and C (22 sites), two campaigns devoted to schools and unexperienced observers. A gaussian fits well with $R^{2}=0.979$.  The shift of its centre toward more luminous stars, $\Delta x =-0.53\pm0.03$ mag, and the smaller dispersion $\sigma=0.50\pm0.03$ mag could be due to the more homogeneous kind of observers, mainly equally unexperienced. The lower panel shows few observations from dataset D, a project devoted to advanced amateurs. Measurements  have been obtained by two experienced observers  searching for the faintest suspected star, with the same atmospheric conditions, stated by contemporary measurements of stellar extinction, and precise geographic positions. They show a small standard deviation ($0.20\pm0.07$ mag) and a mean shifted of  $\Delta x \approx 0.63$  mag toward fainter visible stars in reasonable agreement with Schaefer (1990) formula for expert observers and a 10\% threshold.

Fig.  \ref{calb2} shows the observed versus calculated limiting magnitudes for dataset A (dots), datasets B and C (triangles) and dataset D (open diamonds). The dotted line separates approximately the areas with prominent photopic vision (left) and prominent scotopic vision (right). Solid line correspond to O-C=0 and the dashed lines show deviations of $\pm1$ mag.
Experienced observers of dataset D seems to gain  a magnitude going from photopic to scotopic visibility. The scotopic observations from the other datasets show a larger scatter toward higher magnitudes than photopic observations. If confirmed this could be due to the greater difficulty to observe in scotopic conditions (e.g. the eye sensitivity changes with distance from the center of the retina and it is very low at the center, see Clark 1990). Further observations made in a selected number of sites with different sky brightnesses by a large number of observers in each of them could allow a deeper statistical analysis.

\subsection{Screening}
\label{screening}

In paper I we neglected the presence of mountains which might shield the light emitted from the sources from a fraction of the atmospheric particles along the line-of-sight of the observer.  
Given the vertical extent of the atmosphere in respect to the height of the mountains, the shielding is not negligible only when the source is very near the mountain and both are quite far from the site (Garstang 1989a; see also Cinzano 2000b). 
However when taking into account altitudes in map computation other two effects of screening by terrain elevation can result in plainly visible artefacts, even if mainly within the accuracy estimated for the maps. 
In valleys surrounded by mountains, their lower elevations relative to nearby mountain edges result, when screening is neglected, in a greater sky brightness due to the illumination of the part of the line of sight going from valley altitude to mountain edge altitude. When proper screening from surrounding mountains is taken into account, this part of the line of sight is shielded and does not contribute to sky brightness. Due to the extinction along the longer light path, the sky brightness is lower than at the mountains edges. 
Another effect is produced by small elevations of terrain which can enhance efficiency of the Earth curvature in shielding far sources. This can be evident, e.g. in case of 
 dark areas located at a hundred kilometres from very lighted areas. 

\begin{figure}
\epsfysize14cm 
\hspace{0.4cm}\epsfbox{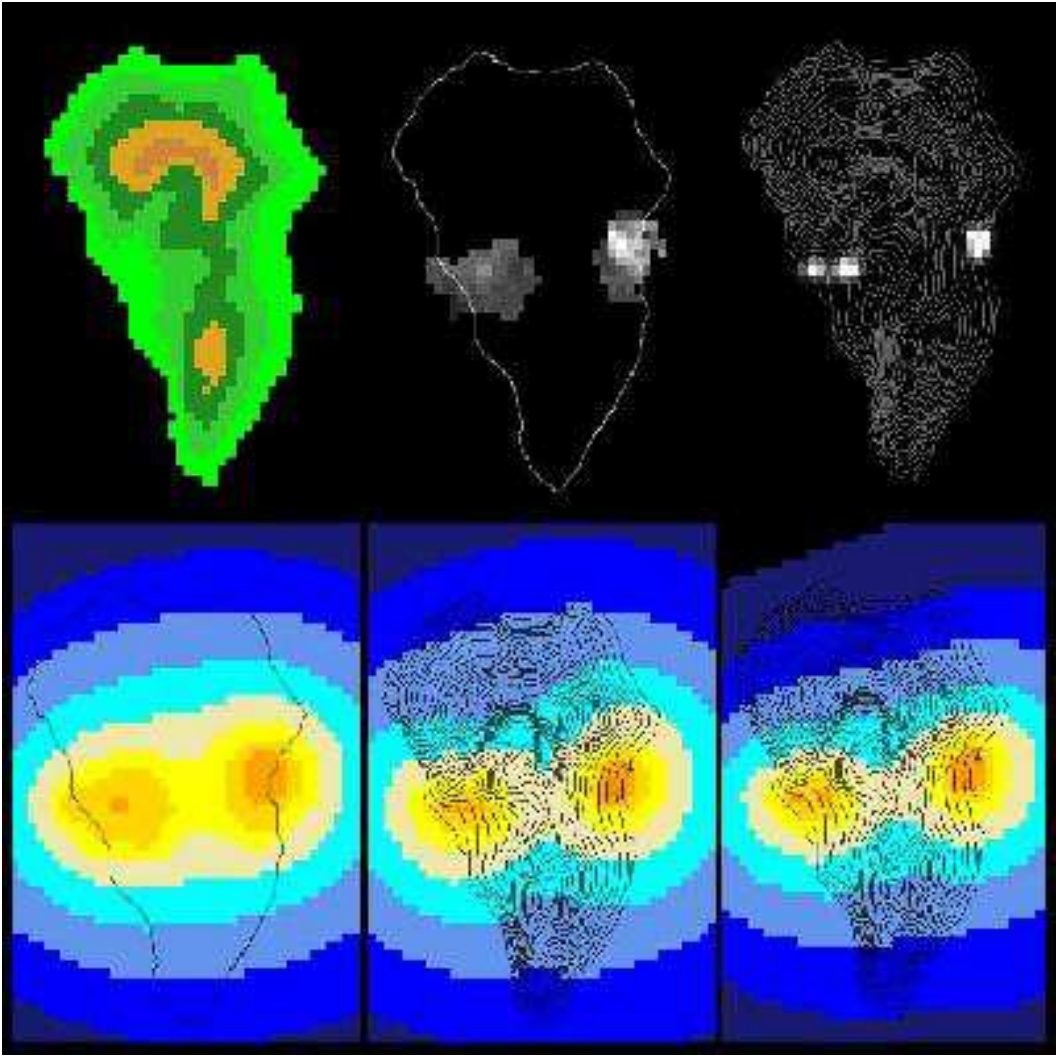} 
\caption[h]{The effects of elevation and screening from nearby mountains at La Palma Island. 
Upper panels show  the terrain elevation (left), the composite satellite image  (middle),  the deconvolved satellite image (right).  Lower panels show the artificial sky brightness at sea level (left) ,  accounting for elevation (middle), accounting for elevation and mountain screening (right). Superimposed are the curves of equal elevation.}
\label{screen1}
\end{figure}

Fig. \ref{screen1}  shows the effects of elevation and screening from nearby mountains at La Palma Island in Canary Island. 
Upper left panel shows  the terrain elevation  in the island (levels indicate elevations of 0-500m, 500-1000m, 1000-1500m, 1500-2000m, $>$2000m), upper middle panel shows the composite radiance calibrated satellite image of the island, the upper right panel show the satellite image after deconvolution. We superimposed the curves of equal elevation. Lower left panel shows the artificial sky brightness at sea level predicted from the original satellite image. 
Lower middle panel shows the artificial sky brightness predicted from the deconvolved data when accounting for elevation. Lower right panel shows the predictions when accounting for elevation and mountain screening. The curves of equal elevation are superimposed. When accounting for elevation, the mountains at North and South of the sources diminish the sky brightness so that the isophotes appear more flattened. However there are no effects where elevation is zero like e.g. in the sea near the top of the image. Screening by mountains has effects also at sea level and the 'umbra' of the mountains is visible looking carefully at the lower right panel. In particular, the screening by the Northern mountains is clearly visible near the top of the map. Colour levels from black to orange in the lower panels indicate zenith artificial brightness of  $<$2.5, 2.5-5.0, 5-10, 10-20, 20-40, 40-80, 80-160, 160-320, 320-640, $>$640 $\mu cd/m^{2}$.  Pollution at zenith is very low at the observatory site Roque de los Muchachos, well under the 10\% of the natural sky brightness over which the International Astronomical Union consider the sky polluted. Maps refers to aerosol content K=1 and do not accounts for local atmospheric conditions and denser aerosol layers. Results for  Canary Island and Chile were not corrected to 1998-1999 and refers to 1996-1997. 

\begin{figure}
\epsfysize6cm 
\hspace{0.4cm}\epsfbox{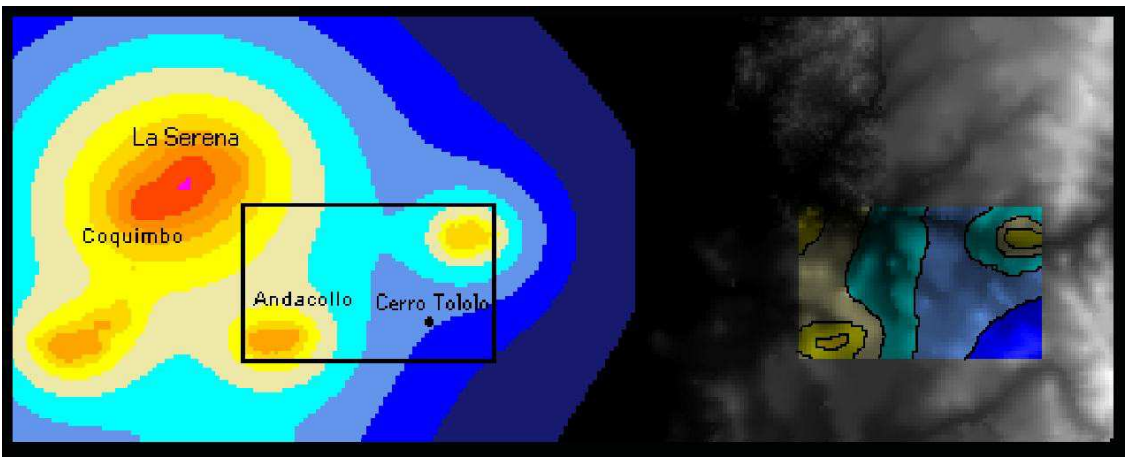} 
\caption[h]{The effects of screening in the nearby of Cerro Tololo. Left panel shows sea level sky brightness. Right panel shows for the selected area the corresponding contours when elevation and screening are accounted. The digital elevation map  is superimposed.}
\label{screen2}
\end{figure}

Fig. \ref{screen2} shows the effect of screening on the sky brightness in the nearby of Cerro Tololo Observatory for the same aerosol content. Colour levels indicate the same artificial brightnesses of fig. \ref{screen1} but we added two more levels. Left panel shows sea level sky brightness neglecting screening and a selected area. Right panel shows for the same area the contours when elevation and screening are accounted for, superimposed to the digital  elevation map. La Serena is about 50 km far from Cerro Tololo so altitude and screening are much less effective than for La Palma. Their effects are recognizable only in the mountains surrounding the other two cities.

Due to the larger computational time requested and to the fact that these effects are in general quite small in respect to the uncertainties, we neglected screening by terrain elevation in the maps of section \ref{s6} . It must be taken into account whenever a greater precision is required and it become a normal practice in future when better code optimisation and faster workstations will be available.

\section{Conclusions}
\label{s7}

We extended the method introduced in Paper I to map the naked eye star visibility and telescopic limiting magnitudes in large territories from DMSP satellite. This requires accounting for altitude of each land area, natural sky brightness in the chosen sky direction, stellar extinction in the chosen photometric band and eye capability. We also  take into account mountain screening for near zenith sky directions.
We presented, as an application,  the maps of naked eye star visibility and total sky brightness in V band in Europe at  zenith.
Maps of limiting magnitudes in  other directions will be useful to predict visibility of astronomical phenomena. A complete mapping of the brightness of the sky at a site, like Cinzano \cite{c99a}, using satellite data instead of population data was already obtained (Cinzano \& Elvidge, in prep.).

Further improvements should be the development of: (i) a faster code to account for screening in a reasonable computational time, (ii) a set of worldwide  three-dimensional atmospheric data sets or models for aerosol and molecules  in order to change from standard clean atmosphere to the typical night atmosphere in each territory in the given season, (iii) a method to measure the upward emission function of each land area from satellite data,  solving problems like the presence of snow or fishing fleets (Cinzano et al., in prep.), (iv) large measurement campaigns in order to better constrain the models (Cinzano \& Falchi 2000). 
We hope that our work will be useful to the comprehension of how much mankind's perception of the Universe is endangered (see also Crawford 1991; Kovalevsky 1992; McNally 1994; Isobe \& 
\newpage
\onecolumn
\twocolumn
\noindent
Hirayama 1998; Sullivan \& Cohen 2000; for a large reference list see Cinzano 1994) and to support the battle against light pollution carried on 
worldwide by the International Dark-Sky Association (see the Web Site www.darksky.org).

\section*{Acknowledgments}

We are indebted to Roy Garstang of JILA-University of  Colorado for his friendly kindness in reading and commenting

on this paper, for his helpful suggestions and for
 interesting discussions.  We acknowledge the EROS Data Center, Sioux Falls, USA for kindly providing us their GTOPO30 digital elevation model and Aragon System for kindly providing us its image analysis program Quantum Image (www.quantumimage.com).

\label{lastpage}
\bsp

\begin{thebibliography}{}

\bibitem[1964]{abra} Abramowitz M., Stegun I.A., 1964, Handbook of Mathematical Functions, Washington, NBS

\bibitem[1973]{allen} Allen C.W., 1973, Astrophysical Quantities, 3rd Ed., Athlone, London

 \bibitem[1976]{berry} Berry R., 1976,  J. R. Astron. Soc. Can., 70, 97-115

\bibitem[1946]{black} Blackwell H. R., 1946, J. Opt. Soc. Am., 36, 624

\bibitem[2000]{cat}  Catanzaro G., Catalano F.A., 2000,  in Cinzano P., ed., Measuring and Modelling Light Pollution, Mem. Soc. Astron. Ital.,  71, 211-220   

\bibitem[1994]{cinref} Cinzano P., 1994, References on Light Pollution and Related Fields v.2, Internal Rep.11 Dep. of Astronomy, Padova, also on-line at {\em www.pd.astro.it/cinzano/refer/index.htm} 



 \bibitem[2000b]{c99a} Cinzano P., 2000a, in Cinzano P., ed., Measuring and Modelling Light Pollution, Mem. Soc. Astron. Ital.,  71,    93-112

 \bibitem[2000c]{c99b} Cinzano P., 2000b, in Cinzano P., ed., Measuring and Modelling Light Pollution, Mem. Soc. Astron. Ital.,   71, 113-130    

 \bibitem[2000d]{c99c} Cinzano P., 2000c, in Cinzano P., ed., Measuring and Modelling Light Pollution, Mem. Soc. Astron. Ital.,   71,  159-166  

\bibitem[2000]{diaz} Cinzano P., Diaz Castro F.J., 2000, in Cinzano P., ed., Measuring and Modelling Light Pollution, Mem. Soc. Astron. Ital.,  71, 251-256   
 
\bibitem[2000]{idaprot} Cinzano P., Falchi F.,  2000, http://www.pd.astro.it/cinzano/ misure/sbeam2.html

\bibitem[2000]{pap1} Cinzano P., Falchi F., Elvidge C.D., Baugh K.E.,  2000a, MNRAS, 318, 641-657 (Paper I)

\bibitem[2000]{cinfal} Cinzano P., Falchi F., Elvidge C.D., Baugh K.E.,  2000b, Mem. Soc. Astron. Ital., accepted

\bibitem[2000]{clark} Clark R.N., 1990, Visual astronomy of the deep sky, Cambridge Univ. Press, New York

\bibitem[2000]{corbo1} Corbo L., 2000a, Astronomia UAI, 2, 45-47

\bibitem[2000]{corbo2} Corbo L., 2000b, http://www.liceorussell.roma.it/Astronomia/ Inquinamento.htm

\bibitem[2000]{corp} Corp L. 1998, http://www.astrosurf.org/mercure/lcorp/pol.htm

\bibitem[2000]{cohen} Cohen J., Sullivan W.T., eds., 2000, The Protection of the Astronomical Sky, UN-IAU Symp. 196, ASP Conf. Ser., in press.

\bibitem[1991]{cra} Crawford D.L., ed., 1991, Light Pollution, Radio Interference and Space Debris, IAU Coll. 112, ASP Conf. Ser.  ~17

\bibitem[1999]{dellap} Della Prugna F. 1999, A\&AS, 140, 345-349
 
\bibitem[1997a]{e97a} Elvidge C.D., Baugh K.E., Kihn E.A., Kroehl H.W., Davis E.R., 1997a, Photogram. Eng. Remote Sens., 63, 727-734

 \bibitem[1997b]{e97b} Elvidge C.D., Baugh K.E., Kihn E.A., Kroehl H.W., Davis E.R., Davis, C., 1997b, Int. J. Remote Sensing, 18, 1373-1379

 \bibitem[1997c]{e97c} Elvidge C.D., Baugh K.E., Hobson V.H., Kihn E.A., Kroehl H.W., Davis E.R., Cocero D., 1997c, Global Change Biology, 3, 387-395

 \bibitem[1999]{e99} Elvidge C.D., Baugh K.E., Dietz J.B., Bland T., Sutton P.C., Kroehl H.W., 1999, Remote Sens. Environ., 68, 77-88

 \bibitem[2000]{e2000} Elvidge C.D., Imhoff, M.L., Baugh, K.E., Hobson, V.R., Nelson, I., Dietz, J.B. 2000, J. Photogrammetry and Remote Sensing, submitted.

\bibitem[1999]{falchi} Falchi F., 1998, Thesis, Univ. Milan

 \bibitem[2000]{falcin} Falchi F., Cinzano P., 2000, in Cinzano P., ed., Measuring and Modelling Light Pollution, Mem. Soc. Astron. Ital.,  71, 139-152    

\bibitem[2000]{fav}  Favero G., Federici A., Blanco A.R., Stagni R., 2000,  in P. Cinzano, ed., Measuring and Modelling Light Pollution, Mem. Soc. Astron. Ital.,  71, 223-230     

 \bibitem[1984]{g84} Garstang R.H., 1984,  Observatory, 104, 196-197 

 \bibitem[1986]{g86} Garstang R.H., 1986,  PASP, 98, 364-375 

 \bibitem[1987]{g87} Garstang R.H., 1987,  in Millis R.L.,  Franz O.G.,  Ables H.D.,  Dahn C.C., eds., Identification, optimization and protection
 of optical observatory sites,  Lowell Obs., Flagstaff,  199-202

 \bibitem[1988]{g88} Garstang R.H., 1988,  Observatory, 108, 159-161 

 \bibitem[1989a]{g89a} Garstang R.H., 1989a,  PASP, 101, 306-329 

 \bibitem[1989b]{g89b} Garstang R.H., 1989b,  ARA\&A, 27,
	    19-40 

 \bibitem[1989c]{g89c} Garstang R.H., 1989c,  BAAS., 21, 2, 759-760 

 \bibitem[1991a]{g91a} Garstang R.H., 1991a,  PASP, 103, 1109-1116  

\bibitem[1991b]{g91b} Garstang R.H., 1991b,  in Crawford D.L., ed., Light Pollution, Radio Interference and Space Debris, IAU Coll. 112,  ASP Conf. Ser.  ~17,
	    56-69

 \bibitem[1991c]{g91c} Garstang R.H., 1991c,  Observatory,
	    111, 239-243 

 \bibitem[1992]{g92} Garstang R.H., 1992, BAAS, 24, 740 

 \bibitem[1993]{g93} Garstang R.H., 1993, BAAS, 25, 1306 

 \bibitem[2000]{g99a} Garstang R.H., 2000a, in Cinzano P., ed., Measuring and Modelling Light Pollution, Mem. Soc. Astron. Ital.,  71, 71-82      

\bibitem[2000]{g99b} Garstang R.H., 2000b, in Cinzano P., ed., Measuring and Modelling Light Pollution, Mem. Soc. Astron. Ital.,  71, 83-92 

 \bibitem[1999]{gc99} Gesch D.B., Verdin, K.L., Greenlee, S.K., 1999, EOS Trans. Am. Geophys. Union, 80, 6, 69-70            

\bibitem[2000]{haenel} Haenel A., 1998, http://www.eso.org/outreach/spec-prog/aol/ market/collaboration/lpoll/

 \bibitem[1998]{isoham98} Isobe S., Hamamura S., 1998, in  Isobe S., Hirayama, T., ed., Preserving the Astronomical
    Windows, Proc. IAU JD5,  ASP Conf. Ser. ~139,  191-199

\bibitem[1998]{iso} Isobe S., Hirayama T., eds., 1998,  Preserving the Astronomical Windows, Proc. IAU JD5, ASP Conf. Ser. ~139

\bibitem[1955]{johnson} Johnson H., 1955, Ann. Ap., 18, 292

 \bibitem[1991]{joseph} Joseph J.H., Kaufman Y.J., Mekler Y., 1991, Appl. Optics, 30, 3047-3058 

\bibitem[1946]{knoll} Knoll H. A., Tousey R., Hulburt E.O., 1946, J. Opt. Soc. Am., 36, 480

\bibitem[1992]{kov} Kovalevsky J., ed., 1992, The Protection of Astronomical and Geophysical Sites, NATO Pilot Study n. 189, Frontieres, Paris 

\bibitem[1987]{ks87} Krisciunas K. 1987, PASP, 99, 887-894

\bibitem[1990]{ks90} Krisciunas K. 1990, PASP, 102, 1052-1063

\bibitem[1997]{ks97} Krisciunas K. 1997, PASP, 109, 1181-1188

\bibitem[1981]{lieske} Lieske R.W., 1981, Proc. of Int. Telemetry Conf., 17, 1013-1020

\bibitem[1978]{mccl} McClatchey R.A., Fenn R.W., Selby J.E.A., Volz F.E., Garing J.S., 1978, in  Driscoll W.G., Vaughan W., eds., Handbook of Optics,  McGraw-Hill, New York

\bibitem[1994]{mcn} McNally D., ed., 1994, The Vanishing Universe, Proc. IAU-ICSU-UNESCO meeting Adverse environmental impacts on astronomy,  Cambridge Univ. Press, Cambridge

\bibitem[2000]{pier}  Piersimoni A., Di Paolantonio A., Brocato E., 2000,  in Cinzano P., ed., Measuring and Modelling Light Pollution, Mem. Soc. Astron. Ital.,  71, 221-222     

\bibitem[2000]{por}  Poretti E., Scardia M., 2000, in Cinzano P., ed., Measuring and Modelling Light Pollution, Mem. Soc. Astron. Ital.,  71, 203-210  

\bibitem[2000]{rm} Roach F.E., Meinel A.B., 1955, ApJ, 122, 554  

\bibitem[1990]{schaefer} Schaefer B. E., 1990, PASP, 102, 212

\bibitem[1993]{schaefer2} Schaefer B. E., 1993, Vistas in Astronomy, 36, 311-361 

\bibitem[1968]{snell} Snell C. M., Heiser A. M., 1968, PASP, 80, 336


 \bibitem[1989]{sul89} Sullivan W.T., 1989,  Int. J. Remote Sensing, 10, 1-5 

 \bibitem[1991]{sul91} Sullivan W.T., 1991, in Crawford D.L., ed., 
    Light Pollution, Radio Interference and Space Debris, IAU Coll. 112,  ASP Conf.
    Ser.  ~17, 11-17

 \bibitem[1973]{tre73} Treanor P.J., 1973,  Observatory, 93,
    117-120 

 \bibitem[1973]{w73} Walker M.F., 1973, PASP, 85, 508-519

 \bibitem[1988]{w88} Walker M.F., 1988, PASP, 100, 496-505

\bibitem[2000]{zit}  Zitelli V., 2000, in  Cinzano P., ed., Measuring and Modelling Light Pollution, Mem. Soc. Astron. Ital.,  71, 193-202

\end{thebibliography}
\end{document}